\documentclass{article}
\usepackage{spconf,amsmath,epsfig,subfigure}
\usepackage{multirow,bigstrut}

\let\OLDthebibliography\thebibliography
\renewcommand\thebibliography[1]{
  \OLDthebibliography{#1}
  \setlength{\parskip}{0pt}
  \setlength{\itemsep}{0pt plus 0.3ex}
}
\usepackage{animate}

\begin{document}\sloppy

\def\x{{\mathbf x}}
\def\L{{\cal L}}

\def \useAnimate {}
\title{An Emerging Coding Paradigm VCM: \\ A Scalable Coding Approach Beyond Feature and Signal
}
%
\name{Sifeng Xia, Kunchangtai Liang, Wenhan Yang, Ling-Yu Duan and Jiaying Liu}
\address{Peking University, Beijing, China}

\maketitle

\begin{abstract}
In this paper, we study a new problem arising from the emerging MPEG standardization effort Video Coding for Machine (VCM)\footnotemark[1],
\footnotetext[1]{https://lists.aau.at/mailman/listinfo/mpeg-vcm}
which aims to bridge the gap between visual feature compression and classical video coding. VCM is committed to address the requirement of compact signal representation for both machine and human vision in a more or less scalable way. To this end, we make endeavors in leveraging the strength of predictive and generative models to support advanced compression techniques for both machine and human vision tasks simultaneously, in which visual features serve as a bridge to connect signal-level and task-level compact representations in a scalable manner. Specifically, we employ a conditional deep generation network to reconstruct video frames with the guidance of learned motion pattern. By learning to extract sparse motion pattern via a predictive model, the network elegantly leverages the feature representation to generate the appearance of to-be-coded frames via a generative model, relying on the appearance of the coded key frames. Meanwhile, the sparse motion pattern is compact and highly effective for high-level vision tasks, \textit{e.g.} action recognition. Experimental results demonstrate that our method yields much better reconstruction quality compared with the traditional video codecs ($0.0063$ gain in SSIM), as well as state-of-the-art action recognition performance over highly compressed videos (9.4\% gain in recognition accuracy), which showcases a promising paradigm of coding signal for both human and machine vision.
\end{abstract}

\begin{keywords}
Video coding for machine,
joint feature and video compression,
human vision,
sparse motion pattern,
frame generation
\end{keywords}


\section{Introduction}
\label{sec:intro}
Video coding aims to compress the videos into a compact form for efficient computing, transmission, and storage.
Many efforts are put into this domain, and over the last three decades, a few coding standards are built to significantly improve the coding efficiency.
The latest video codecs, \textit{i.e.} MPEG-4 AVC/H.264~\cite{Overviewavc} and High Efficiency Video Coding (HEVC)~\cite{Overviewhevc} seek to improve the video coding performance
by edging out spatial, temporal and coding redundancies of video frames.
In the past few years,
data-driven methods have been popular and bring in tremendous progress in the compression task.
The latest data-driven methods have largely overpassed performance of the state-of-the-art codecs, \textit{e.g.} HEVC by further improving various kinds of modules like
intra-prediction~\cite{huintra},
inter-prediction~\cite{fiicip,oneforall},
loop filter~\cite{jiaTIP19,park2016cnn},~\textit{etc}.
These techniques significantly improve the video quality from the perspective of the signal fidelity and human vision.


\begin{figure}[t]
	\centering
	\begin{minipage}{0.23\textwidth}
		\centering
		\ifx \useAnimate \undefined
		\includegraphics[width=0.98\textwidth]{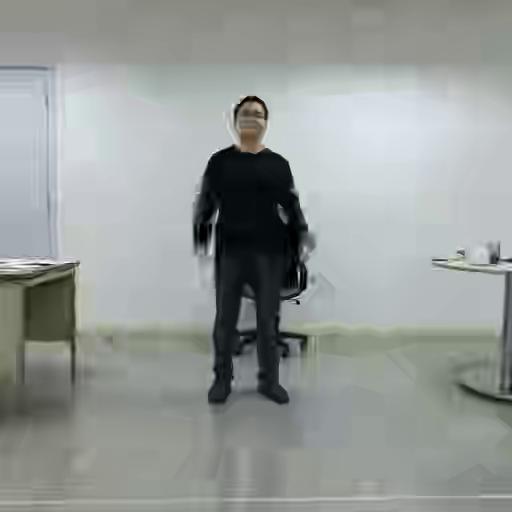}
		\else
		\animategraphics[width=0.98\textwidth,autoplay,loop]{12}{imgs/394_1363_1/hevc_}{5}{20}
		\fi
	\end{minipage}	
	\begin{minipage}{0.23\textwidth}
		\centering
		\ifx \useAnimate \undefined
		\includegraphics[width=0.98\textwidth]{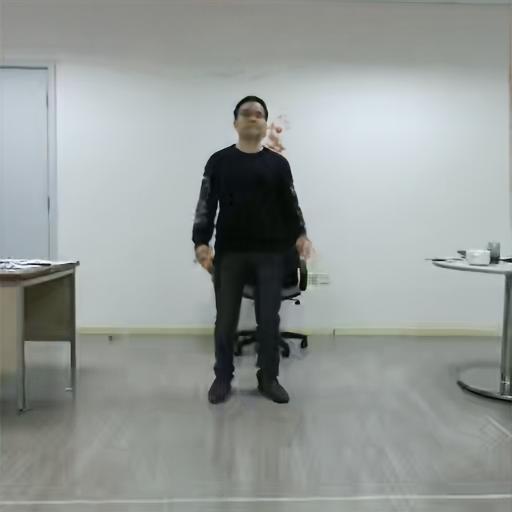}
		\else
		\animategraphics[width=0.98\textwidth,autoplay,loop]{12}{imgs/394_1363_1/pro_}{5}{20}
		\fi
	\end{minipage}
	\caption{
		The visual results of the reconstructed videos by HEVC (left panel) and our method (right panel). Embedded videos are best viewed in Acrobat Reader.
	}
	\label{fig:teaser}
\end{figure}

Existing coding techniques run into problems
when encountering big data and video analytics.
The massive data streaming generated everyday from the smart cities
needs to be compressed, transmitted and analyzed to provide high valuable  information, such as the results of action recognition, event detection, \textit{etc}.
Given this scenario, it is expensive to perform the analysis on the compressed videos,
as the video coding bit-stream is redundant and existing coding mechanism
is not flexible to discard the information that is unrelated to analytical tasks~\cite{rdo_feat}.
Therefore, in the context of big data, it is still an open problem to perform the scalable video coding,
where the requirement of machine vision is first met and additional bitrates can be utilized to further improve visual quality of the reconstructed video progressively and incrementally.
It is an urgent need to obtain a scalable feature representation that connects the information of low and high-level vision
and switches the forms between two purposes freely.

The success of deep learning models has opened a new door.
The deep analytic models can extract compact and high-valuable representations, which can convert the redundant pixel domain information into the sparse feature domain.
In contrast, deep generative models are responsible to produce the whole images and videos with only the guidance of highly abstracted and compact features.
Supported by these tools,
we can realize the scalable compression of videos and features jointly, which is close to both practical application demands in the big data context and accords with the mechanism of human brain circuits.
The most compact and valuable abstracted features are first extracted via deep analytic models~\cite{Zhu_lstm,Song_attention,Liu_pku_mmd} to support the analytics applications.
With these features, we can locate the place and time where some key events happen, namely rethinking rough situations.
Then, guided by the features, other information is partly generated by deep generative models~\cite{gan,PSGAN,chan2019dance,Siarohin_2019_CVPR}, and partly compressed and decoded to support the video reconstruction, namely rethinking scene details.
This solution is potential to address the difficulty in combining video analytics and reconstruction in the big data streaming, which is the main target of video coding for machine (VCM).
The first step of the process can provide timely analytical results with a small portion of bitrates to fulfill the need of machine vision and  the second stage can further provide the reconstructed videos with regards to the analytical results using more bitrates to meet the need of human vision.

\begin{figure*}[t]
	\centering
	\includegraphics[width=18cm]{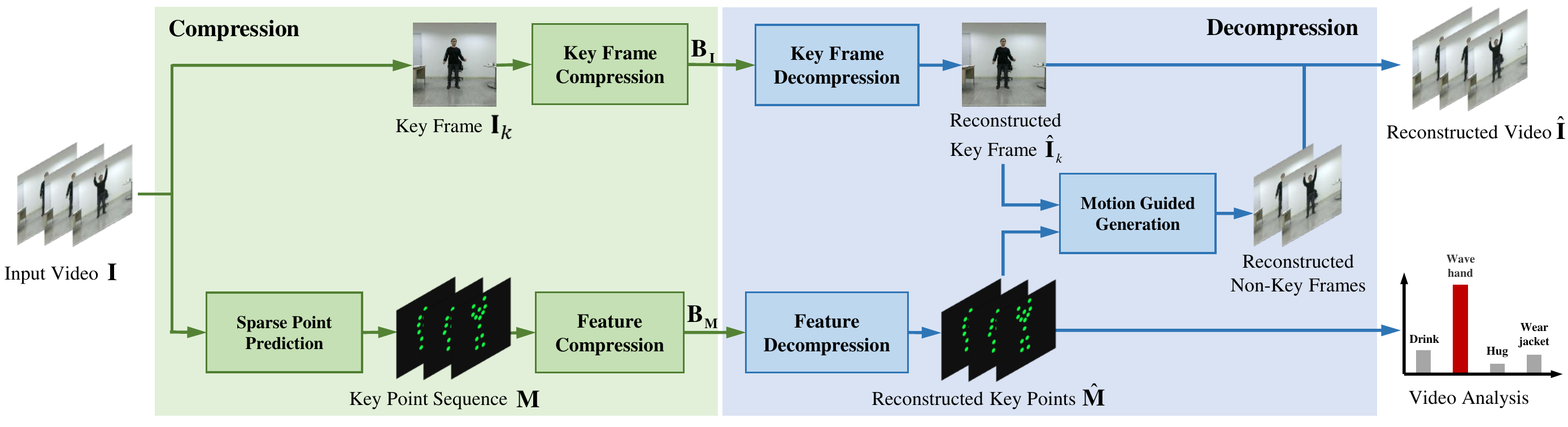}
	\caption{
		The coding pipeline of our proposed joint feature and video compression that serves for both human and machine vision.
	}
	\label{fig:f1}
\vspace{-5mm}
\end{figure*}

Specifically,
in this paper, we propose a scalable joint compression method for both features and videos in surveillance scenes, where a learnable motion pattern bridges the gap between machine and human vision.
The sparse motion pattern is first extracted automatically via a deep predictive model.
After that, the appearance of the currently coded frame is transfered from the coded key frame with the guidance of the motion pattern via a deep generative model.
The sparse motion pattern is highly efficient for high-level vision tasks, \textit{e.g.} action recognition, and it can also meet the requirement of human vision.
In this way, the total coding cost of features and videos can be largely reduced.

In summary, the contributions of our paper are summarized as follows:
\begin{itemize}
	\item To the best of our knowledge, we make the first attempt towards VCM to compress features and videos jointly, serving for both machine and human vision. A novel scalable compression framework is designed with the aid of predictive and generative models to support both machine and human vision.
\vspace{-2mm}
	\item In our framework, the learned sparse motion pattern is used as a bridge, which is flexible and largely reduces the total coding cost of two kinds of vision. To promote the analysis performance of human action recognition, we additionally apply the constraint of the learned points with the guidance of human skeletons.
\vspace{-2mm}
	\item  Compared with traditional video codecs, our method not only achieves much better video quality but also offers significantly better action recognition performance at very low bitrates, which showcases a promising paradigm of coding signal for both human and machine vision.
\end{itemize}

The rest of the article is organized as follows. Sec. \ref{sec:pipline} illustrates the pipeline of our proposed joint feature and video compression. The detailed network architecture for key point prediction and motion guided target video generation is also elaborated. Experimental results are shown in Sec. \ref{sec:exp} and concluding remarks are given in Sec. \ref{sec:conclu}.

\vspace{-3mm}
\section{Joint Compression of Features and \\ Videos}
\label{sec:pipline}
Given a video sequence ${\bf{I}} =\{ {{I_1},{I_2},...,{I_N}} \}$ where $N$ indicates the frame number, it is necessary to compress ${\bf{I}}$ for transmission and storage.
In this section, we will first analyze limitations of traditional video coding methods.
Then, we develop our new framework to compress features and videos jointly in a scalable way.

\vspace{-3mm}
\subsection{Sequential Compression and Analytics}

The traditional video codec targets to optimize the visual quality of the compressed video from the perspective of signal fidelity.
In this process, all frames are coded.
For each frame, spatial and temporal predictions are utilized to predict the target frame with existing coded signal to remove the spatial and temporal redundancy.
Then, the prediction residue and much syntax information are coded for reconstruction at the decoder side.
Though the data can be efficiently compressed via the latest codecs, the scale of data is still massive as a huge amount of data is taken all days and weeks.
Therefore, it is intractable to compress and save data with a high quality, and analyze it later.

It is a reasonable trade-off to compress the data into a low-quality format.
However, existing compression methods which target at
optimizing the human vision are not desirable for high-level analytics tasks.
If we lower the quality of the compressed videos, the performance of action recognition will be largely degraded. As demonstrated in Sec. \ref{actionacc}, our method uses only about 1/3 bitrate cost of the traditional compression method to achieve a better performance in the action recognition task.
Another path that leads to effective video analytics is to extract and compress features.
However, in this case, we could not obtain the reconstructed videos.
This also sets barriers to real applications,
where the results usually need to be confirmed by human examiners.
Therefore, we seek to develop a flexible and scalable framework which compresses the feature at first for machine vision and reconstructs the video later for human vision with more bits consumption.

\vspace{-3mm}
\subsection{An Overview of Joint Feature and Video Compression}
Fig.~\ref{fig:f1} has illustrated the overview pipeline of
the proposed joint feature and video compression method.
The motivation lies in the fact that in surveillance scenes,
the videos can be represented as a background layer (static or slow moving) and
moving objects, such as human bodies.
Then, the network is capable of learning to represent a video sequence with
the learned sparse motion pattern, which can indicate the object motion among frames.
In our work, we focus on indoor surveillance videos with a static background and moving humans.

At the encoder side, with the captured video frames ${\bf{I}} = \{ {{I_1},{I_2},...,{I_N}} \}$, a set of key frames ${{\bf{I}}_k}$ will be first selected and compressed with traditional video codecs and form the bit-stream ${{\bf{B}}_{\bf{I}}}$.
The coded key frames convey the appearance information which includes the background and human appearances and will be transmitted to the decoder side to synthesize the non-key frames.
Moreover, the learned Sparse Point Prediction Network (SPPN) extracts sparse key points from video frames and form a point sequence ${\bf{M}} = \{ {{m_1},{m_2},...,{m_N}} \}$.
The sparse point sequence can mark the motion areas in the frames and convey the motion trajectories of objects along the temporal dimension, which is viewed as a sparse motion pattern of the video.
The point sequence will also be coded to a bit stream ${{\bf{B}}_{\bf{M}}}$ for transmission.

At the decoder side, key frames will be first reconstructed from ${{\bf{B}}_{\bf{I}}}$ and we indicate the reconstructed key frames as ${{{\bf{\hat I}}}_k}$.
For reconstructing remaining non-key frames, the key points are decompressed as ${\bf{\hat M}} = \{ {{{\hat m}_1},{{\hat m}_2},...,{{\hat m}_N}} \}$ and a learned Motion Guided Generation Network~(MGGN) will first estimate the motion flow among frames based on the decompressed sparse motion pattern.
Then, MGGN transfers the appearance of the reconstructed key frames to remaining non-key frames with the guidance of the estimated motion flow.
Specifically, for the $t$-th frame to be reconstructed, we denote its previous key frame as ${{{\hat I}_{k}}}$. The target frame is synthesized as ${{\hat I}_t} = \varphi ( {{{\hat I}_k},{{\hat m}_k},{{\hat m}_t}} )$, where $\varphi$ represents MGGN. Finally, the reconstructed key points ${{\bf{\hat M}}}$ and the video ${\bf{\hat I}} = \{ {{{\hat I}_1},{{\hat I}_2},...,{{\hat I}_N}} \}$ can be used respectively for machine analysis and human vision.

\vspace{-3mm}
\subsection{Detailed Network Architecture Illustration}
\label{sec:generation}
The critical feature of our joint feature and video compression framework is to be capable of capturing the motion between video frames for both machine analytics and video reconstruction.
There are several kinds of ways to model video motion, such as dense optical flow \cite{PWCNet} or sparse motion representations based on human poses \cite{chan2019dance} or unsupervisely learned key points \cite{Siarohin_2019_CVPR}.
In our work, we hope the motion representations to be sparse enough for efficient machine analytics.
Therefore, we refer to \cite{Siarohin_2019_CVPR} to predict key points of frames as the sparse motion pattern, which is compact enough that costs only a few bits for transmission and storage. For human vision, motion flow among video frames will be later derived from the sparse motion pattern to guide the generation of the target frame.

\begin{figure}[t!]
	\centering
	\includegraphics[width=8.3cm]{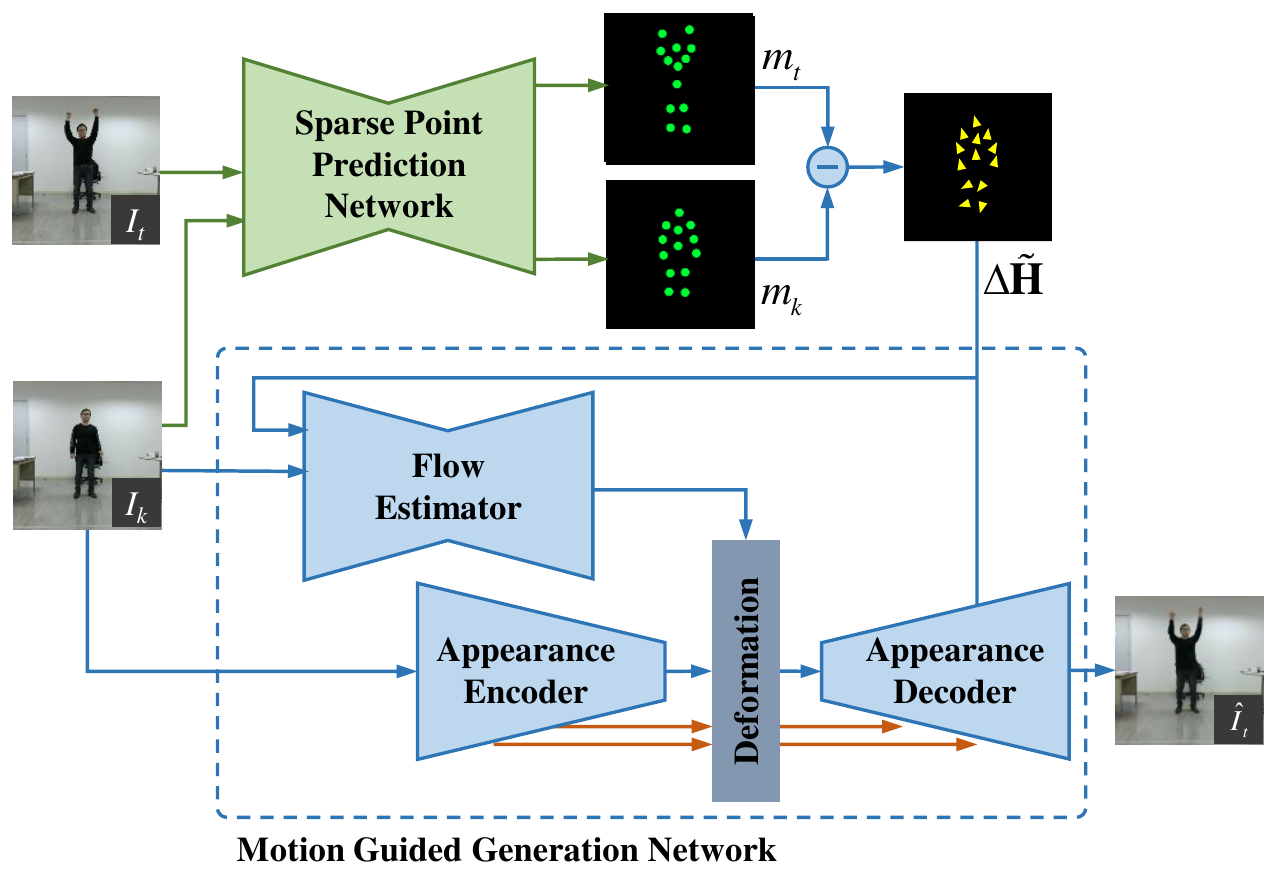}
	\caption{
		Framework of our proposed joint feature and video compression,
		including
		a sparse point prediction network and
		motion guided generation network to
		extract the sparse motion pattern and
		generate the target frame.
	}
	\label{fig:f2}
\vspace{-5mm}
\end{figure}

The framework of the network is shown in Fig. \ref{fig:f2}. For a key frame $I_k$ and a target frame $I_t$ which is to be generated at the decoder side, their key points will be first predicted by SPPN, and this sparse motion pattern is later combined with $I_k$ for estimating the flow map between frames.
Then, the generated flow map will guide the transfer of the appearance of $I_k$ to the target frame. Details of different parts of the network are described as follows.

\noindent \textbf{Sparse Point Prediction.}
For an input frame, a sub-network of the U-Net architecture followed by softmax activations is used to extract $L$ heatmaps ${\bf{H}} = \left\{ {{H_1},...{H_L}} \right\}$ for key point prediction. Each heatmap ${H_l} \in {\left[ {0,1} \right]^{{\rm{H \times W}}}}$ corresponds to one key point position $p_l$, which is estimated as follows:
\begin{equation}
\label{eq1}
{p_l} = \sum\limits_{p \in \Omega } {{H_l}\left[ p \right]p},
\end{equation}
where $\Omega$ is the set of positions of all pixels. Besides the key point position, the corresponding covariance matrix ${\Sigma _l}$ is defined as:
\begin{equation}
\label{eq2}
{\Sigma _l} = \sum\limits_{p \in \Omega } {{H_l}\left[ p \right]\left( {p - {p_l}} \right){{\left( {p - {p_l}} \right)}^{\rm{T}}}}.
\end{equation}
The covariance matrix is generated here because it can additionally capture the correlations between the key point and its neighbor pixels.
Consequently, for each key point, totally 6 float numbers including two numbers indicating the position and 4 numbers in the covariance matrix are used for description.

For the succeeding usage, the key point description will be used to generate new heatmaps by a Gaussian-like function.
This operation is done for that the new heatmaps are more compatible with convolutional operations. Specifically, the new heatmap  ${{\tilde H}_l}$ will be generated as follows:
\begin{equation}
\label{eq3}
{{\tilde H}_l}\left[ p \right] = \exp \left( { - \alpha {{\left( {p - {p_l}} \right)}^{\rm{T}}}\Sigma _l^{ - 1}\left( {p - {p_l}} \right)} \right),
\end{equation}
where $\alpha$ is a normalization constant and set to $0.5$. After this progress, two sets of newly generated heatmaps ${{{\bf{\tilde H}}}^k} = \left\{ {\tilde H_1^k,...,\tilde H_L^k} \right\}$ and ${{{\bf{\tilde H}}}^t} = \left\{ {\tilde H_1^t,...,\tilde H_L^t} \right\}$ are generated from frames $I_k$ and $I_t$, respectively.
\vspace{2mm}

\noindent \textbf{Motion Flow Estimation}.
With the estimated key points and newly generated heatmaps, a sub-network in MGGN will be first used to estimate the motion flow between frames $I_k$ and $I_t$. The source frame $I_k$ is adopted to form the input for it conveys the appearance information. Meanwhile, the difference heatmaps $\Delta {\bf{\tilde H}} = {{{\bf{\tilde H}}}^t} - {{{\bf{\tilde H}}}^k}$ between two frames are used to form the input to provide sparse motion information. The flow estimator will finally output a flow map ${\xi _{k \to t}}$.

\noindent \textbf{Motion Guided Target Frame Generation}.
The target frame is generated with a sub-network of the U-Net architecture. Feature maps of different sizes are extracted by the appearance encoder and will be bypassed to the appearance decoder for feature fusion. In order to align the features to the target frame, features will be previously deformed with the estimated flow map ${\xi _{k \to t}}$ before fusion. Besides, the difference heatmaps $\Delta {\bf{\tilde H}} = {{{\bf{\tilde H}}}^t} - {{{\bf{\tilde H}}}^k}$ is used as side information that is inputted to the appearance decoder. Then, the target frame ${{\hat I}_t}$ can be generated by the appearance decoder.

\noindent \textbf{Skeleton Guided Point Prediction Loss Function}. In \cite{Siarohin_2019_CVPR}, the key points prediction is learned unsupervisely. In our work, we additionally use human skeleton information to guide the key point prediction.
The skeleton information is used for its high efficiency in modeling human actions as the skeleton points cover many human joints, which are highly correlated to human actions.
Consequently, the PKU-MMD dataset \cite{pkummd} is used in our work for training and testing, which is a large-scale dataset and contains many human action videos.
More importantly, human skeletons are available in this dataset for each human body in the videos.

We sample 16 skeleton points for each human body and employ an $L_1$ loss function for supervision. The key point detection loss function is defined as follows:
\begin{equation}
\label{eq4}
{{\cal L}_{{\rm{point}}}} = \frac{1}{n}\sum\limits_{i = 1}^n {\sum\limits_{l = 1}^{16} {\parallel {p_l^i - \pi _l^i} \parallel_1} } ,
\end{equation}
where $\pi _l^i$ represents the $l$-th skeleton point of the human in the $i$-th training sample.

\noindent \textbf{Overall Loss Function}. Besides the point prediction loss, a combination of an adversarial and the feature matching loss proposed in \cite{ganloss} are used for training. The discriminator $D$ will take ${{{\bf{\tilde H}}}^t}$ concatenated with either the real image $I_t$ or the generated image ${{\hat I}_t}$ as its input. The discriminator and generator losses are calculated as follows:
\begin{equation}
\label{dloss}
{{\cal L}_D} = {{\mathop{\rm E}\nolimits} _{{I_t}}}[ {{{( {D( {{I_t},{{{\bf{\tilde H}}}^t}} ) - 1} })^2}} ] + {E_{( {{I_t},{{\hat I}_t}} )}}[ {{{( {D( {{{\hat I}_t},{{{\bf{\tilde H}}}^t}} )} )}^2}} ] ,
\end{equation}
\vspace{-5mm}
\begin{equation}
\label{gloss}
{{\cal L}_G} = {E_{({I_t},{{\hat I}_t})}}[{(D({{\hat I}_t},{{{\bf{\tilde H}}}^t}) - 1)^2}].
\end{equation}

For a better reconstruction quality, a reconstruction loss function ${\cal L}_{rec}$ is built to keep $I_t$ and ${\hat I}_t$ to have similar feature representations. ${\cal L}_{rec}$ is implemented by calculating the $L_1$ distance between features extracted from $I_t$ and ${\hat I}_t$ by the discriminator. Features outputted by all layers of the discriminator are all used for calculation.

The final loss function is calculated by ${\cal L} = {\lambda _{po{\mathop{\rm int}} }}{{\cal L}_{{\rm{point}}}} + {\lambda _{rec}}{{\cal L}_{{\rm{rec}}}} + {{\cal L}_G}$, where ${\lambda _{po{\mathop{\rm int}} }}$ and ${\lambda _{rec}}$ are respectively set to $20$ and $10$.

\section{Experiments}
\label{sec:exp}
\subsection{Experimental Details}
PKU-MMD dataset \cite{pkummd} is used to generate the training and testing samples.
In total $3317$ clips with $32$ frames are sampled for training and $227$ clips with $32$ frames are sampled for testing. All frames are cropped and resized to $512\times512$ during sampling. The skeleton information is also used during the training process. 16 skeleton points are chosen for each frame and mapped to the corresponding two-dimensional space to generate the labels for key point prediction. The network is implemented in PyTorch and the Adam optimizer \cite{adam} is used for training. We randomly select two frames from a clip to form a training sample.

In the testing process, we consistently use the first frame in each clip as the key frame. At the encoder side, the key frame is coded with the HEVC codec in the constant rate factor mode. The constant rate factor is set to $32$. Besides the key frame, key points of all frames in the clip are predicted by SPPN
and compressed for transmission. As mentioned in Sec. \ref{sec:generation}, each key point contains 6 float numbers. For the two position numbers, a quantization with the step $2$ is performed for compression.
For the other 4 float numbers belonging to the covariance matrix, we calculate the inverse of the matrix in advance, and then quantize the 4 values with a step $64$.
Then, the quantized key point values are further losslessly compressed by the Lempel Ziv Markov chain algorithm (LZMA) algorithm \cite{lzma}.
At the decoder side, the compressed key frame and points are decompressed and used to generate remaining frames.

To verify the efficiency of our coding paradigm, we use HEVC as the anchor for comparison by additionally compressing all frames with the HEVC codec. The constant rate factor is firstly consistently set to 51, the highest compression ratio.
Then, the recognition accuracies of using the learned sparse motion pattern and the compressed videos are compared.
To verify the reconstruction quality, we set the constant rate factor to 44 and compare the reconstruction results between HEVC and our method with similar coding cost. The reconstruction quality is compared both quantitatively and qualitatively.

\vspace{-3mm}
\subsection{Action Recognition Accuracy}
\label{actionacc}
We identify the efficiency of the learned key points for high-level analytics tasks in the action recognition task. Although there are 6 numbers for each key point, we only use two quantized position numbers for action recognition. Consequently, only bits of the compressed position numbers are considered for calculating the bitrate cost of feature-based action recognition. To align to the bitrate cost of the features, we firstly resize all clips to the size of $256*256$ and then use the constant rate factor $51$ to compress the testing clips with HEVC.

 \begin{figure*}[t]
	\centering
\subfigure[Ground Truth]{
	\includegraphics[width=0.15\linewidth]{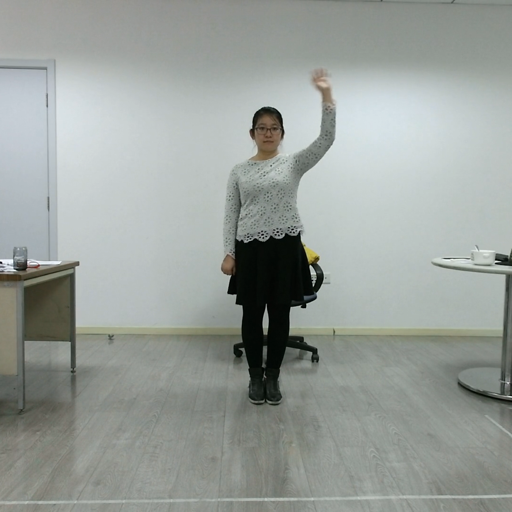}
	\includegraphics[width=0.15\linewidth]{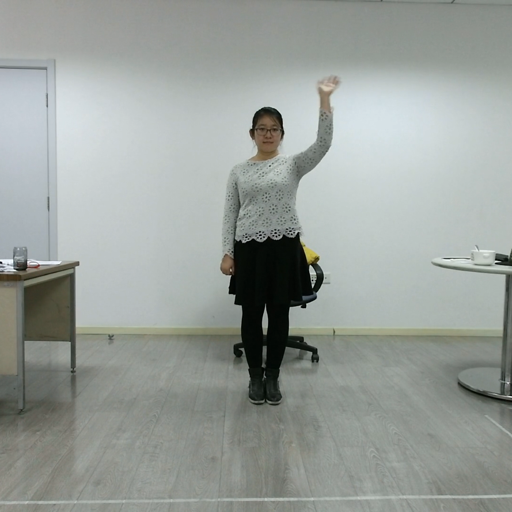}
    \includegraphics[width=0.15\linewidth]{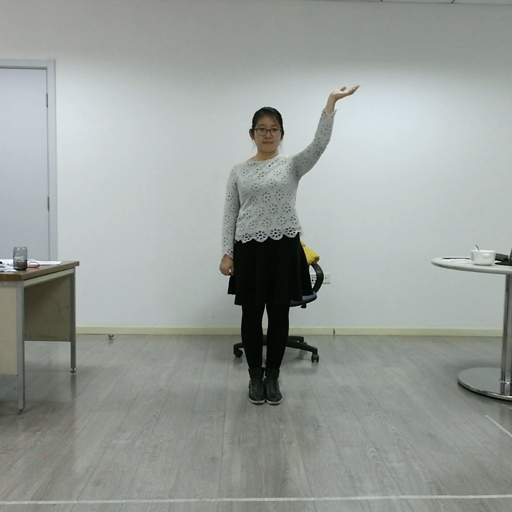}
    \hspace{3mm}
    \includegraphics[width=0.15\linewidth]{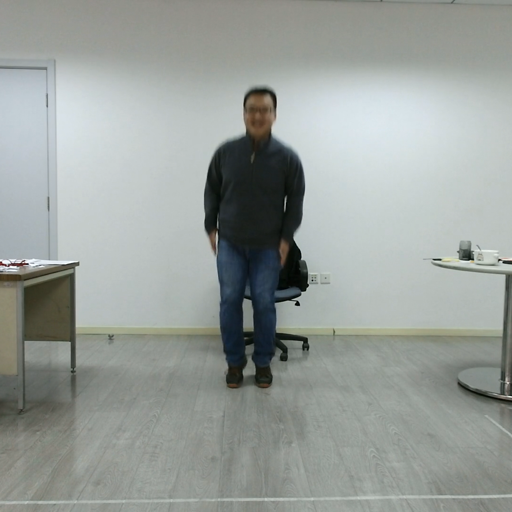}
    \includegraphics[width=0.15\linewidth]{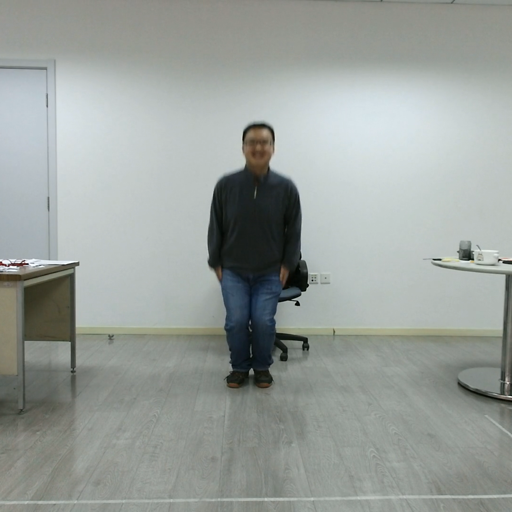}
    \includegraphics[width=0.15\linewidth]{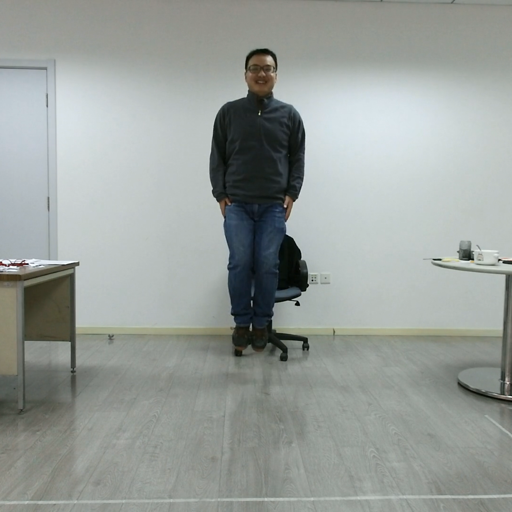}
    }
\subfigure[HEVC]{
	\includegraphics[width=0.15\linewidth]{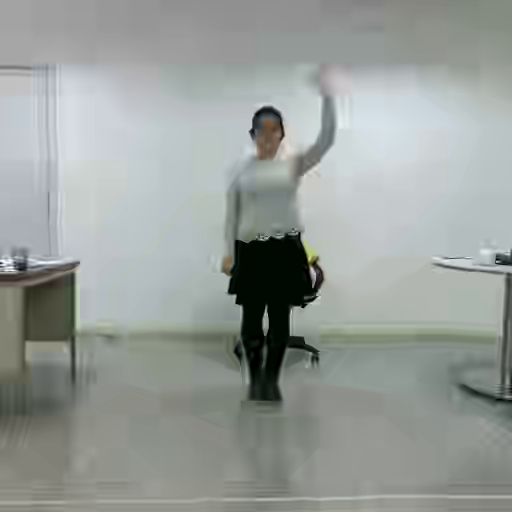}
	\includegraphics[width=0.15\linewidth]{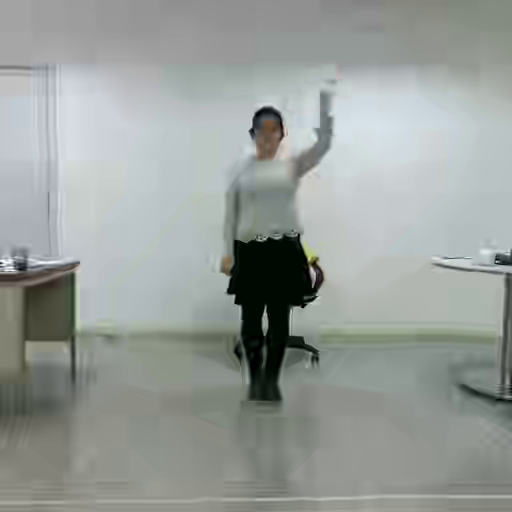}
    \includegraphics[width=0.15\linewidth]{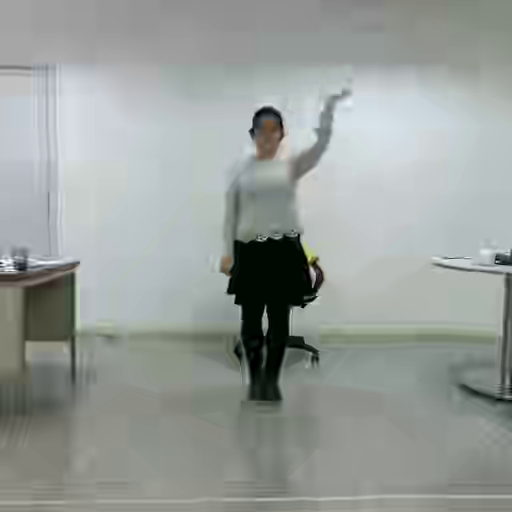}
      \hspace{3mm}
    \includegraphics[width=0.15\linewidth]{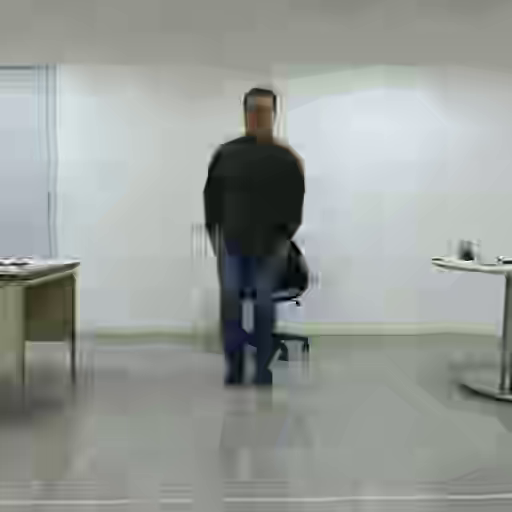}
    \includegraphics[width=0.15\linewidth]{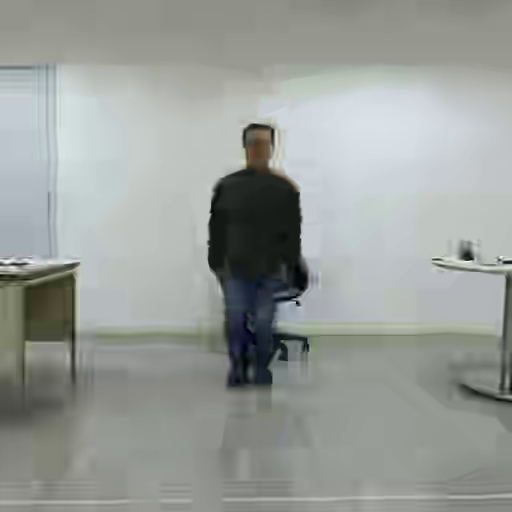}
    \includegraphics[width=0.15\linewidth]{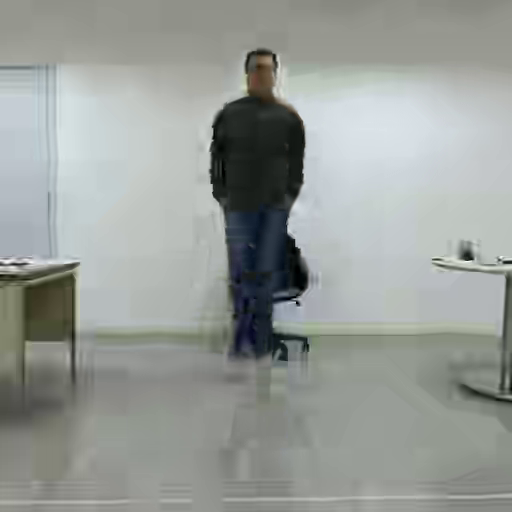}
    }
\subfigure[Proposed]{
	\includegraphics[width=0.15\linewidth]{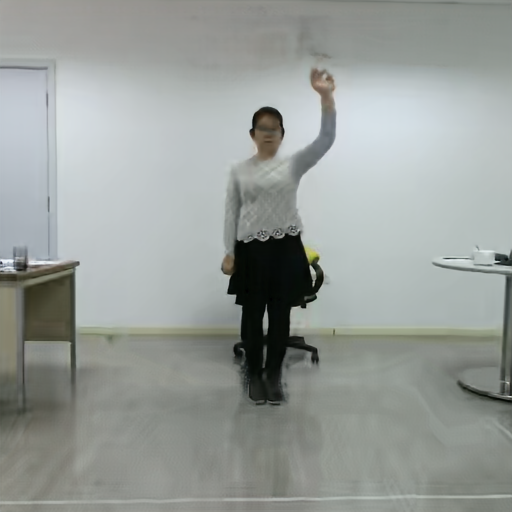}
	\includegraphics[width=0.15\linewidth]{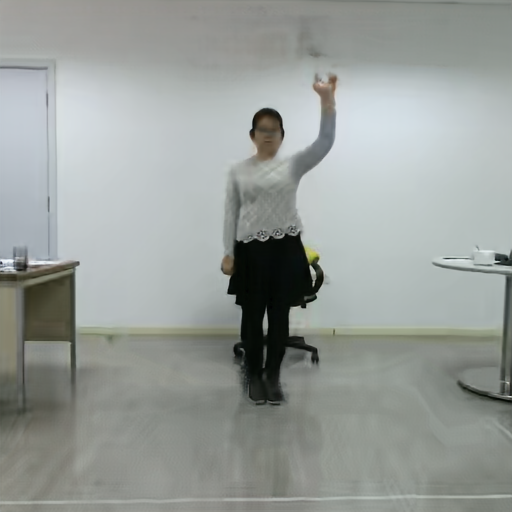}
    \includegraphics[width=0.15\linewidth]{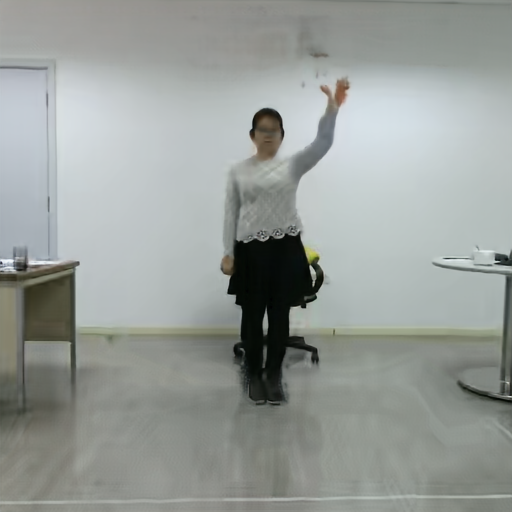}
      \hspace{3mm}
    \includegraphics[width=0.15\linewidth]{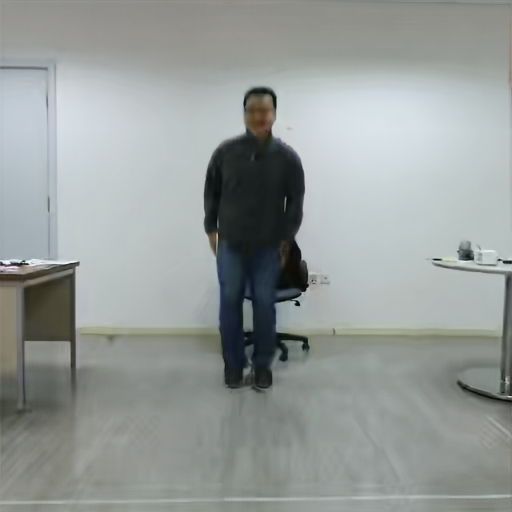}
    \includegraphics[width=0.15\linewidth]{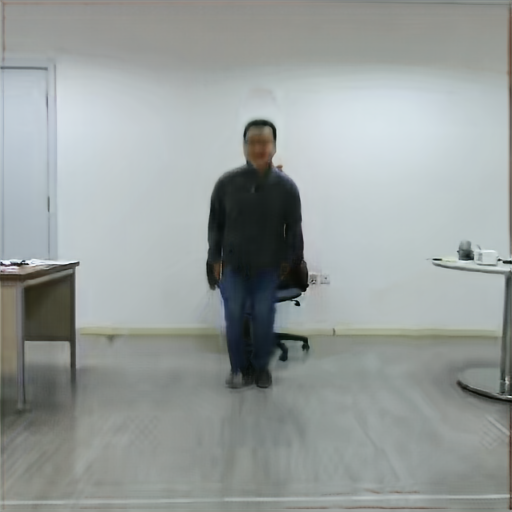}
    \includegraphics[width=0.15\linewidth]{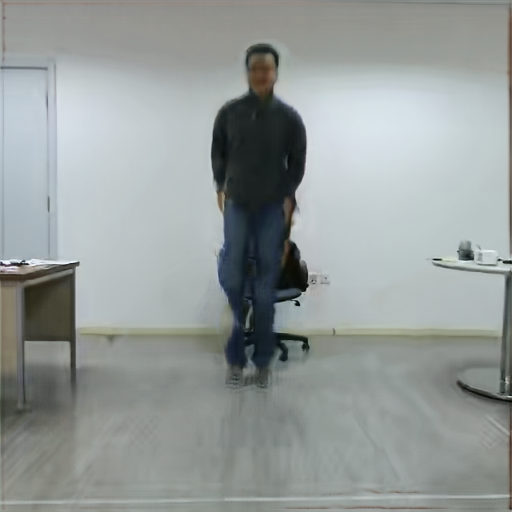}
    }
    \vspace{-3mm}
\caption{Video reconstruction results of different methods.
Left and right three panels correspond to two video clips in the testing set, respectively.
The average SSIM values of the reconstructed clips are respectively $0.8889$ and $0.9204$ for HEVC and the proposed method for the left clip. For the right clip, the SSIM values of HEVC and the proposed method are $0.8966$ and $0.9143$.}
	\label{fig:f3}
    \vspace{-3mm}

\end{figure*}

\begin{table}[htb]
  \centering
\caption{Action recognition accuracy of different methods and corresponding bitrate costs.}
 \begin{tabular}{c|c|c}
 \hline
  Input     & Bitrate (Kbps) & Accuracy(\%) \bigstrut\\
 \hline
 \hline
Compressed Video & 16.2  & 65.2  \bigstrut\\
Compressed Key Point & 5.2   & 74.6  \bigstrut\\
 \hline
 \end{tabular}
 \label{tab1}
 \end{table}

 Table \ref{tab1} has shown the action recognition accuracy and corresponding bitrate costs of different kinds of data. Our method can obtain considerable action recognition accuracy with only $5.2$ Kbps bitrate cost. Although we have chosen the worst coding quality, it still needs $16.2$ Kbps to transform and store the compressed videos. More bitrates cannot bring too much performance improvement in action recognition on compressed videos. Unfortunately, the recognition accuracy even drops by $9.4\%$.
\vspace{-2mm}

\begin{table}[t]
    \centering
	\caption{SSIM comparison between different methods and corresponding bitrate costs.}
	 \label{tab2}
	 \begin{tabular}{c|c|c}
	 \hline
	    Codec & Bitrate (Kbps) & SSIM \bigstrut\\
	 \hline
	 \hline
	 HEVC  & 33.0  & 0.9008  \bigstrut\\
	 Ours  & 32.1  & 0.9071  \bigstrut\\
	 \hline
	 \end{tabular}%
	 \vspace{-3mm}
\end{table}

\subsection{Video Reconstruction Quality}
The video reconstruction quality of the proposed method is also compared with that of HEVC. During the testing phase, we compress the key frames with the constant rate factor $32$ to maintain a high appearance quality.
The bitrate is calculated by jointly considering the compressed key frames and key points.
As for HEVC, we compress all frames with the constant rate factor $44$ to achieve an approaching bitrate cost.

Table \ref{tab2} has shown the quantitative reconstruction quality of different methods. SSIM values are adopted for quantitative comparison. It can be observed that, our method can achieve better reconstruction quality than HEVC with a fewer bitrate cost. Subjective results of different methods are shown in Fig. \ref{fig:f3}. There are obvious compression artifacts on the reconstruction results of HEVC, which heavily degrade the visual quality. Compared with HEVC, our method can provide far more visually pleasing results.

\section{Conclusion}
\label{sec:conclu}
\vspace{-3mm}
In our work, we propose a novel framework to bridge the gap between compression for features and videos.
A conditional deep generation network is designed to
reconstruct video frames with the guidance of a learned sparse motion pattern.
This representation is highly compact and also effective for high-level vision tasks, \textit{e.g.} action recognition.
Therefore, it is scalable to meet the requirements of both machine and human vision, which reduces the total coding cost.
Experimental results demonstrate that our method can obtain superior reconstruction quality and action recognition accuracy with fewer bitrate costs compared with traditional video codecs.

\vspace{-3mm}
\small
\bibliographystyle{IEEEbib}
\bibliography{refs}

\end{document}